# MHD stability of fully non-inductive discharges in TORE SUPRA

by P. Maget, F. Imbeaux, G. Huysmans, Ph. Moreau, M. Ottaviani, J.-L. Segui
Association Euratom-CEA, CEA Cadarache, F-13108 Saint-Paul-lez-Durance, France

Fully non-inductive discharges are performed on the Tore Supra tokamak using Lower Hybrid (LH) waves in a current drive scheme. The MHD aspect is a crucial issue for these discharges, since after hitting an MHD limit it may run into a regime where a strong permanent MHD activity develops. This so-called MHD regime [Maget04] degrades fast electron confinement as well as LHCD efficiency and electron confinement in the plasma core. In the 2003 campaign, a milestone has been reached with the injection of more than 1 GJ of energy in the tokamak, in stationary discharges lasting more than 6 minutes. But the MHD regime has been encountered in several discharges, while the record discharges have shown a very stationary (m=3, n=2) MHD activity in the plasma core without triggering the MHD regime (see figure 1).

With the exploration of several global parameters such as the toroidal magnetic field (B), the parallel refractive index of LH waves ($n_{//}$) and the total plasma current ($I_p$), a map of MHD properties for fully non-inductive discharges has been drawn. This is achieved by extrapolating in $I_p$ reference magnetic equilibriums from stationary pulses, while imposing a constant shape of pressure and current profiles, and a poloidal beta $\beta_p \sim 1/I_p$. Note that in the experimental conditions explored, the LH deposition profile is globally peaked, but hollow in the very core, and the region of reversed magnetic shear extends while $n_{//}$ increases.

Linear stability properties are investigated with the CASTOR code [Kerner89], which gives the linear growth rate of resistive MHD modes in toroidal geometry. They show the existence of stable domains for $q_{min}>2$ and for $1.5>q_{min}>1$, in agreement with the analysis of past experiments [Wijnands94, Litaudon96, Zabiego01]. But most of the explored domain is MHD unstable, including the point where the record pulses have been performed with a remarkable reliability (see figure 2). Saturated tearing modes are indeed observed in several cases without triggering the deleterious MHD regime.

Considering the non-linear evolution of MHD modes is necessary for discrimination between harmless saturated modes and the MHD regime. A particular pulse where a slow ramp down of the total plasma current was performed up to the triggering of the MHD regime allows to identify the mechanism of transition. It can be attributed to the full reconnection of the double tearing mode, which flattens the electron temperature profile up to the outer resonant surface. Its condition of occurrence can be estimated from the calculation of the helical flux relative to the resonant surface $q=m/n$: $\psi^* = \int d\psi (m/n - q)$. The full reconnection happens when $\psi^*$ is close to zero at the outer resonant surface [Carreras79]. The domain where the MHD regime can be triggered is therefore approximately given by the condition $m/n > q_{min} > q(\psi^*_{outer} = 0)$. Such an evaluation is consistent with the experimental points where the MHD regime is indeed triggered (see figure 3), although a safe interval exists for $q_{min}$ slightly below a rational (as for record pulses), while full reconnection occurs for $\psi^*_{outer}$ slightly above zero, in agreement with non-linear MHD simulations. Note that the regime where full reconnection is observed is tearing-like, scaling with resistivity as $\eta^{3/5}$, and is consistent with the 'intermediate regime' described by Ishii et al [Ishii00].

As a result, a map of MHD properties for fully non-inductive discharges is established, which can be used for the design of future experiments. For longer-term purpose, the drawing of such a map could be implemented in a real-time module for self-programmed discharges, which could avoid deleterious MHD activity while optimizing the discharge performance.

References:

[Maget04] Maget P. *et al* 2004 *Nucl. Fusion* **44** 443
[Kerner89] Kerner W. *et al* 1989 *J. Comput. Phys.* **85** 1
[Wijnands94] Wijnands T. *et al* 1994 *Nucl. Fusion* **37** 777
[Litaudon96] Litaudon X. *et al* 1996 *Plasma Phys. Control. Fusion* **38** 1603
[Zabiego01] Zabiego M. *et al* 2001 *Nucl. Fusion* **43** 1625
[Carreras79] Carreras B. *et al* 1979 *Nucl. Fusion* **19** 583
[Ishii00] Ishii Y. *et al* 2000 *Phys. Plasmas* **7** 4477

Figures:

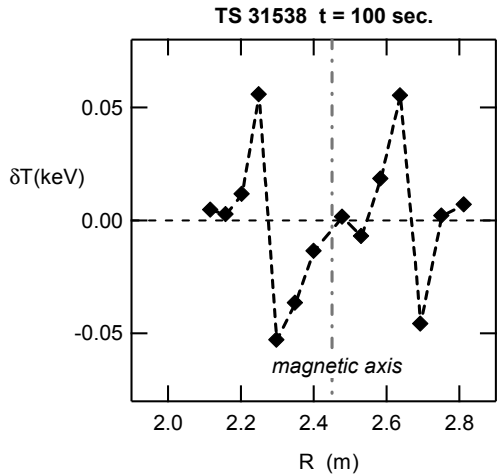

*Figure 1: Radial structure of electron temperature fluctuations after 100 seconds of plasma for pulse TS-31538*

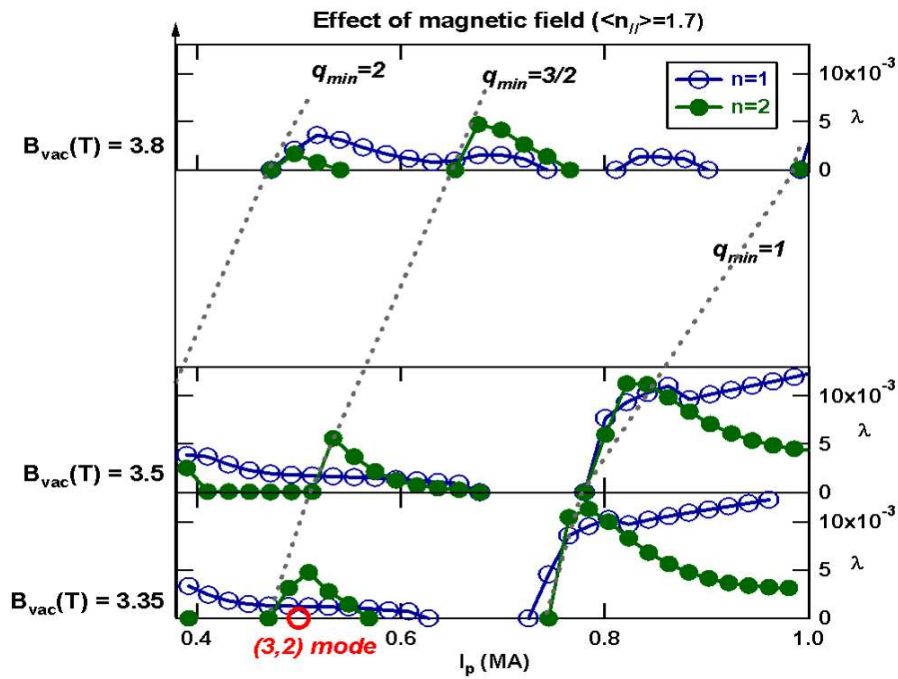

*Figure 2: Linear growth rate ($\lambda=\gamma\tau_A$) of n=1 and n=2 resistive MHD modes, as a function of the total plasma current for fully non-inductive discharges realized at 3 different magnetic*

fields. At B=3.35T is indicated the position where record pulses have been performed, with the observation of a very stationary (m=3,n=2) mode.

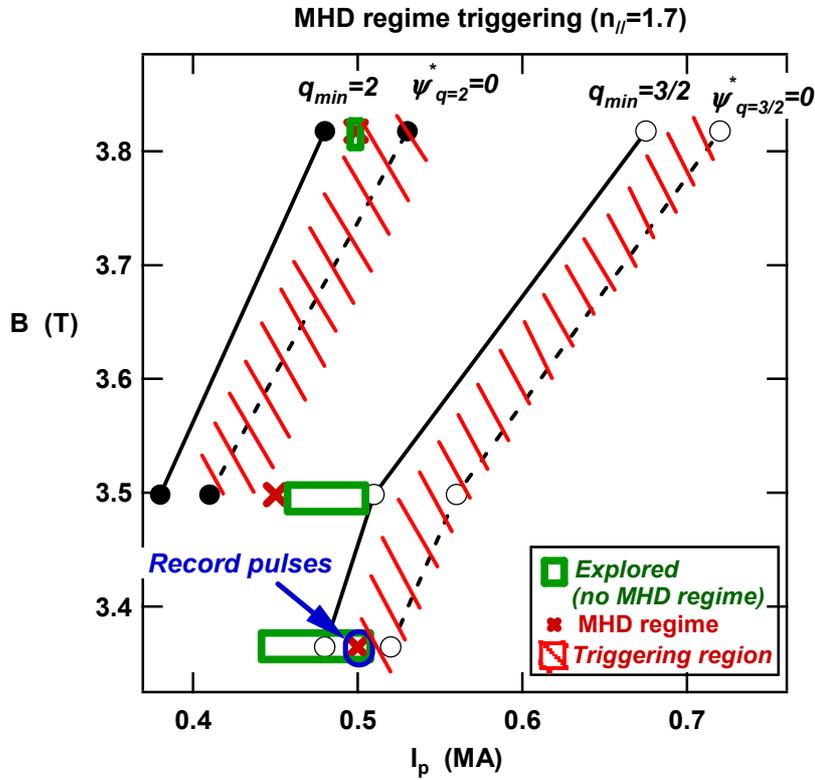

Figure 3: Domain (B, $I_p$) where MHD regime can be triggered, for $n_{//}$=1.7. Explored regions and observed transitions are indicated. Frontiers of rational $q_{min}$ and $\psi^*_{outer}$ =0 are obtained from reference pulses. The triggering region is in fact for $q_{min}$ slightly below a rational (as for record pulses), down to a q-profile where $\psi^*$ is slighly above zero at the outer resonant surface.